\documentclass[twocolumn,prb]{revtex4}
\usepackage{amsfonts}
\usepackage[T1]{fontenc}
\usepackage{amsmath,amsbsy,amssymb,graphicx}
\usepackage{times}

\begin{document}

\title{Topological Toda lattice and nonlinear bulk-edge correspondence}
\author{Motohiko Ezawa}
\affiliation{Department of Applied Physics, University of Tokyo, Hongo 7-3-1, 113-8656,
Japan}

\begin{abstract}
The Toda lattice is a model of nonlinear wave equations allowing exact
soliton solutions. It is realized by an electric circuit made of a
transmission line with variable capacitance diodes and inductors. It has
been generalized to the dimerized Toda lattice by introducing alternating
bondings specified by a certain parameter $\lambda $, where it is reduced to
the Toda lattice at $\lambda =0$. In this work, we investigate the
topological dynamics of the voltage along the transmission line. It is
demonstrated numerically that the system is topological for $\lambda <0$,
while it is trivial for $\lambda >0$ with the phase transition point given
by the original Toda lattice ($\lambda =0$). These topological behaviors are
well explained by the chiral index familiar in the Su-Schrieffer-Heeger
model. The topological phase transition is observable by a significant
difference between the dynamics of voltages in the two phases, which is
explained by the emergence of the topological edge states. This is a
bulk-edge correspondence in nonlinear systems. The dimerized Toda lattice is
adiabatically connected to a linear system, which would be the reason why
the topological arguments are valid. Furthermore, we show that the
topological edge state is robust against randomness.
\end{abstract}

\maketitle

\textbf{Introduction}: Topological insulators is one of the most intensively
studied topics in condensed matter physics\cite{Hasan,Qi}. The emergence of
topological edge states is characteristic to a topological phase, as is the
bulk-edge correspondence. They are conveniently realized in electric circuits%
\cite%
{TECNature,ComPhys,Hel,Lu,YLi,EzawaTEC,Research,Zhao,EzawaLCR,EzawaSkin,Garcia,Hofmann,EzawaMajo,Tjunc,Lee}%
. It is based on the fact that we can construct such a circuit Laplacian
that is equivalent to the Hamiltonian. Thus, a circuit Laplacian may have
topological phases. Here, topological edge states are observed by impedance
peaks. There is also an application to dynamical systems. A quantum walk may
be described by the time evolution of the voltage in an electric circuit,
because the Kirchhoff law is rewritten in the form of the Scr\"{o}dinger
equation\cite{QWalk,EzawaUniv,EzawaDirac}. Here, the emergence of a
topological edge state is observable by a peculiar behavior of a quantum
walker, i.e., by a peculiar time evolution of the voltage. The topological
physics has so far been mainly studied in linear systems. One of the merit
of electric circuits is that we can naturally introduce nonlinear elements
into circuits. It is an interesting problem to study topological properties
in nonlinear systems\cite{Kot,Sone} .

The Toda lattice is one of the most famous exactly solvable nonlinear models%
\cite{Toda,Toda2,Toda3}. The Toda lattice with alternating bonding is called
the dimerized Toda lattice\cite{Kofane,Pelap}, where soliton solutions seem
to be absent\cite{Lus}. This is not surprising because the Toda soliton is
nontopological and fragile against perturbations. It is known that the Toda
lattice is realized by a transmission line consisting of variable
capacitance diodes and inductors\cite%
{Hirota,Singer,Yemele,Yemele2,Pelap05,Houwe}. When the inductance is
alternating, it is generalized to the dimerized Toda lattice\cite%
{Kofane,Pelap}. So far there is no study on topological properties of the
Toda lattice.

In this paper, we explore the topological aspect of the dynamics governed by
the nonlinear wave equation associated with the dimerized Toda lattice.
Interestingly we find that the nonlinear wave equation contains a term
proportional to the Su-Schrieffer-Heeger (SSH) Hamiltonian. We wonder if the
dynamics is characterized by the topological number inherent to the SSH
model. The main issue of this work is to study this problem. We solve the
time evolution of the voltage along the transmission line numerically in
nonlinear models, by applying a voltage only at its edge as an initial
condition. We demonstrate that the dynamical behaviors are quite distinct in
the topological and trivial phases assigned by a topological number defined
in the SSH Hamiltonian. The voltage shows a standing wave at the edge in the
topological phase but it rapidly decreases in the trivial phase. These
distinct behaviors are interpreted by the emergence of the topological edge
states, namely, by the bulk-edge correspondence in nonlinear systems. This
would be due to the fact that the nonlinear system is adiabatically
connected to the linear-wave system described by the SSH model. The
topological phase transition point corresponds to the original Toda lattice,
which is unchanged by the strength of nonlinearity. Finally, we check
numerically that the topological edge states are robust against randomness
in inductance.

\begin{figure}[t]
\centerline{\includegraphics[width=0.48\textwidth]{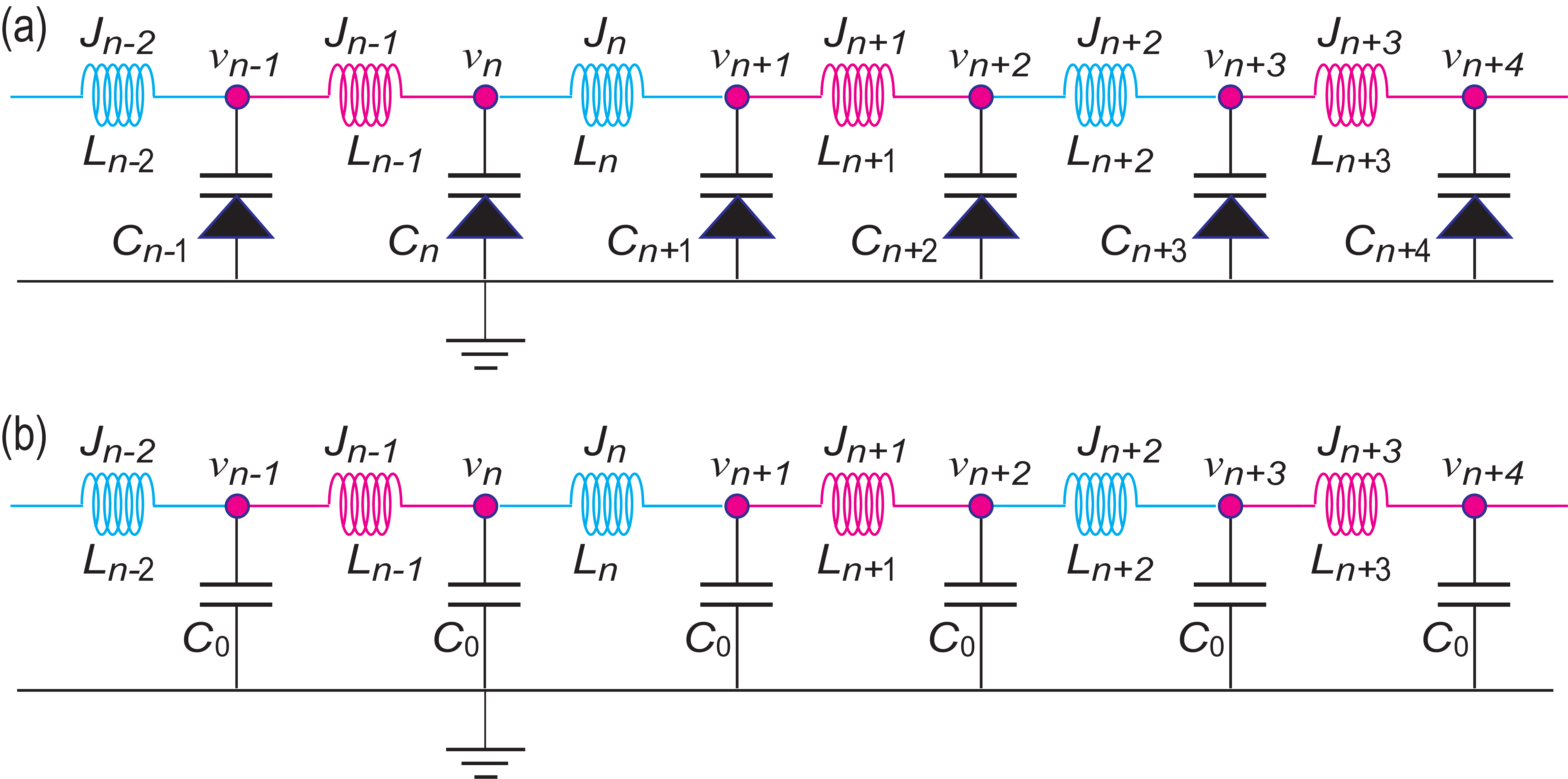}}
\caption{Illustration of a transmission line made of (a) nonlinear elements
realizing the site-dependent Toda lattice and (b) linear elements realizing
the telegrapher equation. The inductance is alternating, as indicated by
magenta and cyan colors. Each node is grounded via a variable capacitance
diode and a capacitor in (a) and (b), respectively.}
\label{FigIllust}
\end{figure}

\textbf{Dimerized Toda Lattice}: The Toda equation\cite{Toda,Toda2} is well
realized by a transmission line with variable capacitance diodes and
inductors\cite{Hirota} as shown in Fig.\ref{FigIllust}(a). The Kirchhoff law
is described by 
\begin{equation}
L_{n}\frac{dJ_{n}}{dt}=v_{n}-v_{n+1},\quad \quad \frac{dQ_{n}}{dt}%
=J_{n-1}-J_{n},
\end{equation}%
where $v_{n}$ is the voltage, $J_{n}$ is the current and $Q_{n}$ is the
charge at the node $n$, while $L_{n}$ is the inductance for the inductor
between the nodes $n$ and $n+1$, as illustrated in Fig.\ref{FigIllust}(a).
The Kirchhoff law is summarized in the form of the second-order differential
equation, 
\begin{equation}
\frac{d^{2}Q_{n}}{dt^{2}}=\frac{1}{L_{n-1}}\left( V_{n-1}-V_{n}\right) -%
\frac{1}{L_{n}}\left( V_{n}-V_{n+1}\right) ,  \label{EqC}
\end{equation}%
where we have introduced a new variable $V_{n}$ by $v_{n}=V_{0}+V_{n}$. The
capacitance is a function of the voltage $V_{n}$\ in the variable
capacitance diode, and it is well given by $C\left( V_{n}\right) =Q\left(
V_{0}\right) /\left( F_{0}+V_{n}-V_{0}\right) $, where $F_{0}$ is a constant
characteristic to the variable capacitance diode\cite{Nakajima}.

\begin{figure}[t]
\centerline{\includegraphics[width=0.48\textwidth]{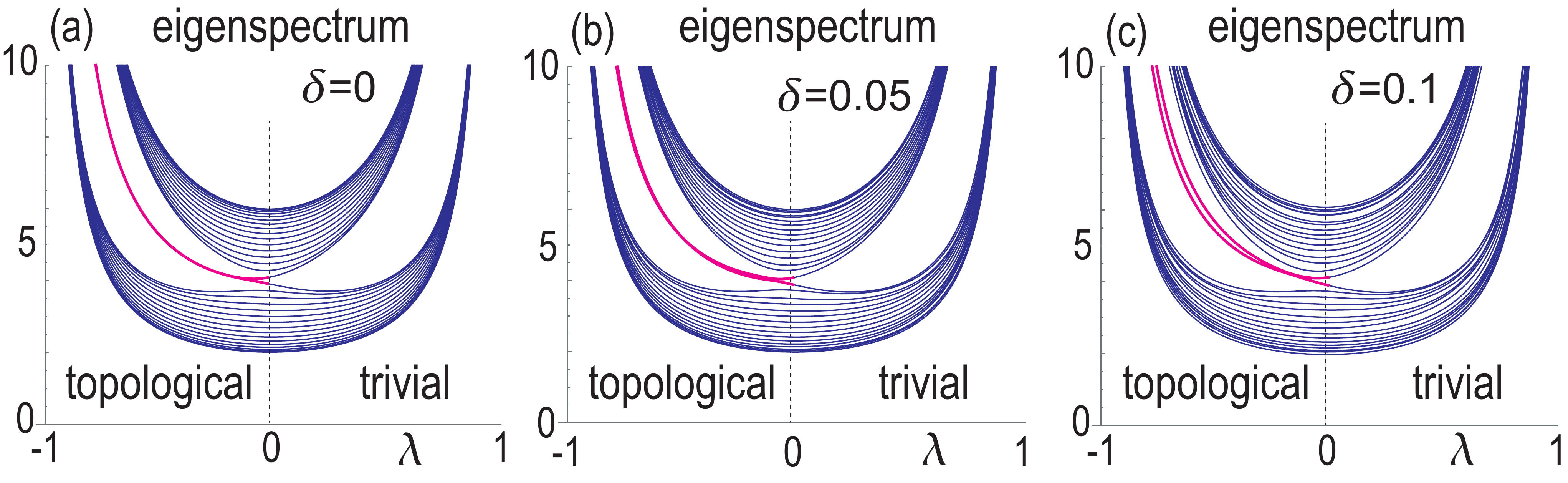}}
\caption{(a)$\sim $(c) Eigenspectrum of $M_{nm}$ as a function of $\protect%
\lambda $. The edge states emerge for $\protect\lambda <0$, as shown by
magenta curves. (a) Pure system ($\protect\delta =0$). (b) and (c) System
with randomness $(\protect\delta \not=0)$. }
\label{FigEdge}
\end{figure}

\begin{figure*}[t]
\centerline{\includegraphics[width=0.98\textwidth]{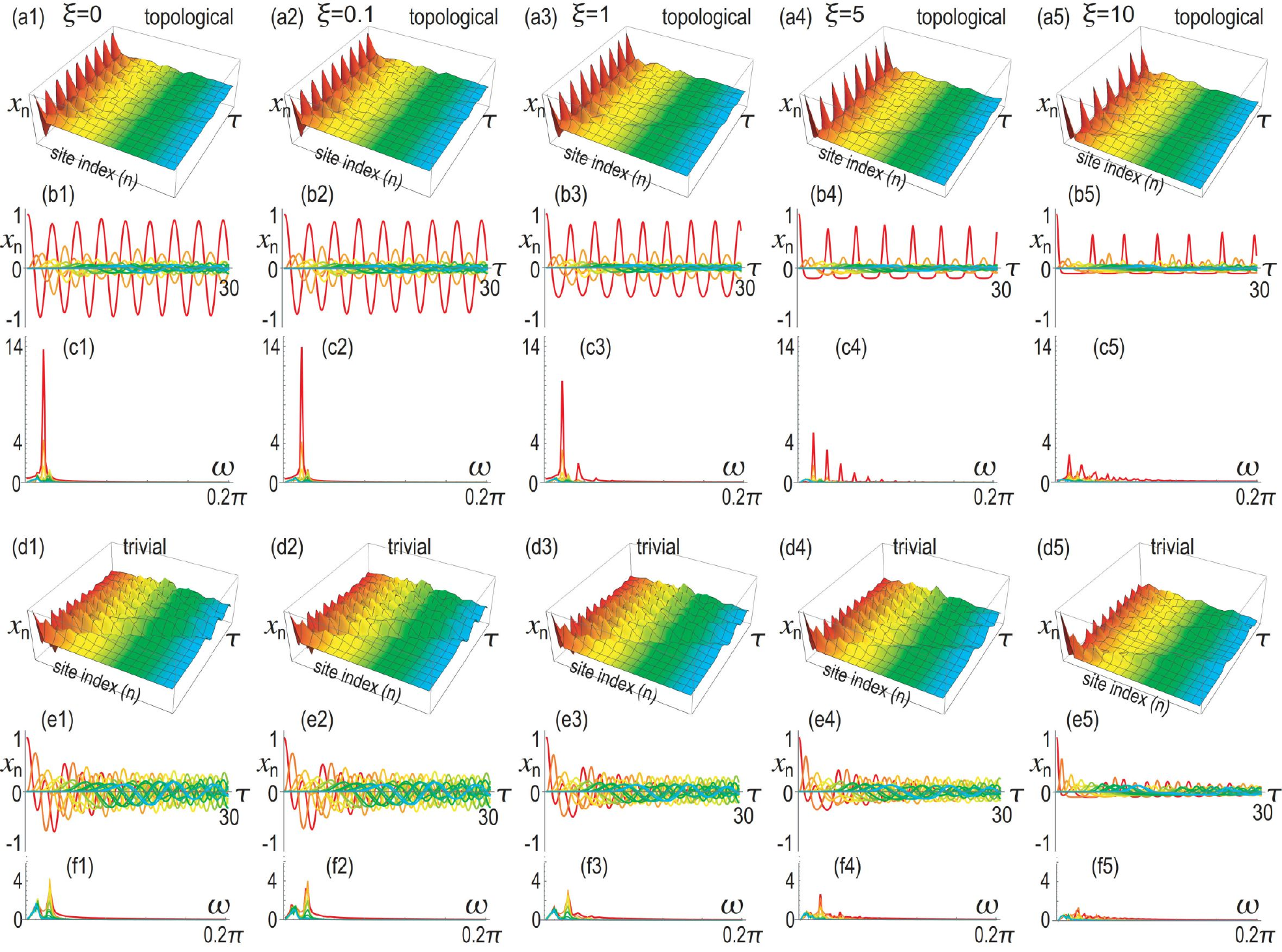}}
\caption{Time evolution of the dimensionless voltage $x_{n}$\ in the
nonlinear telegrapher equation. The horizontal axis is the dimensionless
time $\protect\tau $ ranging $0\leq \protect\tau \leq 30$. (a1)$\sim $(c5)
Topological phase ($\protect\lambda =-0.5$). (d1)$\sim $(f5) Trivial phase ($%
\protect\lambda =0.5$). (a1)$\sim $(a5) and (d1)$\sim $(d5) Bird's eye's
view of the time evolution. (b1)$\sim $(b5) and (e1)$\sim $(e5) Color plot
of the time evolution. (c1)$\sim $(c5) and (f1)$\sim $(f5) Fourier
components in the frequency $\protect\omega $. The red (cyan) curve
represents the voltage at the left (right) edge. We have set $\protect\xi =0$
for (a1)$\sim $(f1), $\protect\xi =0$ for (a2)$\sim $(f2), $\protect\xi =1$
for (a3)$\sim $(f3) , $\protect\xi =5$ for (a4)$\sim $(f4) and $\protect\xi %
=10$ for (a5)$\sim $(f5). }
\label{FigDynamics}
\end{figure*}

The charge is given by%
\begin{equation}
Q_{n}=\int_{0}^{V_{n}}C\left( V\right) dV=Q\left( V_{0}\right) \log \left[
1+V_{n}F_{0}^{-1}\right] +\text{const.}  \label{EqD}
\end{equation}%
A closed form of the differential equation for $V_{n}$ follows from Eqs.(\ref%
{EqC}) and (\ref{EqD}), 
\begin{align}
Q\left( V_{0}\right) \frac{d^{2}}{dt^{2}}& \log \left[ 1+V_{n}F_{0}^{-1}%
\right]  \notag \\
& =\frac{V_{n+1}}{L_{n}}-\left( \frac{1}{L_{n-1}}+\frac{1}{L_{n}}\right)
V_{n}+\frac{V_{n-1}}{L_{n-1}}.  \label{EqA}
\end{align}%
When we set $L_{n}=L$ for all $n$, it is reduced to%
\begin{equation}
Q\left( V_{0}\right) \frac{d^{2}}{dt^{2}}\log \left[ 1+V_{n}F_{0}^{-1}\right]
=\frac{1}{L}\left( V_{n+1}-2V_{n}+V_{n-1}\right) .  \label{TeleToda}
\end{equation}%
This is the Toda equation.

We focus on the dimerized Toda lattice defined by setting $L_{2n-1}=L-\ell $%
, $L_{2n}=L+\ell $, as corresponds to the inductance alternating, where $%
\left\vert \ell \right\vert <L$. We rewrite Eq.(\ref{EqA}) as%
\begin{align}
\frac{1}{\xi }\frac{d^{2}}{d\tau ^{2}}& \log \left[ 1+\xi x_{2n-1}\right] 
\notag \\
& =t_{A}x_{2n}-\left( t_{A}+t_{B}\right) x_{2n-1}+t_{B}x_{2n-2},  \label{EqE}
\\
\frac{1}{\xi }\frac{d^{2}}{d\tau ^{2}}& \log \left[ 1+\xi x_{2n}\right] 
\notag \\
& =t_{B}x_{2n+1}-\left( t_{A}+t_{B}\right) x_{2n}+t_{A}x_{2n-1},  \label{EqF}
\end{align}%
where we have introduced dimensionless quantities,%
\begin{eqnarray}
\tau &=&t/\sqrt{LQ\left( V_{0}\right) V_{1}^{2}/F_{0}},\quad
x_{n}=V_{n}/V_{1},  \notag \\
\xi &=&V_{1}/F_{0},\quad \lambda =\ell /L.
\end{eqnarray}%
They are dimensionless time, voltage, nonlinearity parameter and
dimerization parameter, respectively. We have defined $t_{A}=1/(1-\lambda )$
and $t_{B}=1/\left( 1+\lambda \right) $, where $A$ and $B$ denote the
sublattice indices. Eqs.(\ref{EqE}) and (\ref{EqF}) are summarize as%
\begin{equation}
\frac{1}{\xi }\frac{d^{2}}{dt^{2}}\left( \log \left[ 1+\xi x_{n}\right]
\right) =M_{nm}x_{m},  \label{TeleEq}
\end{equation}%
with a tridiagonal matrix $M_{nm}$, where the nonlinearity is controlled by $%
\xi $.

We make a Fourier transformation from the node index $n$ to the momentum $k$
by setting $x_{A}\left( k\right) =N^{-1/2}\sum_{n=1}x_{2n-1}e^{ink}$ and $%
x_{B}\left( k\right) =N^{-1/2}\sum_{n=1}x_{2n}e^{ink}$, to find that%
\begin{align}
\frac{1}{\xi }\frac{d^{2}}{d\tau ^{2}}& \left( 
\begin{array}{c}
\log \left[ 1+\xi \sum_{k}x_{A}\left( k\right) e^{-ink}\right] \\ 
\log \left[ 1+\xi \sum_{k}x_{B}\left( k\right) e^{-ink}\right]%
\end{array}%
\right)  \notag \\
& \qquad \qquad =\sum_{k}M\left( k\right) \left( 
\begin{array}{c}
x_{A}\left( k\right) \\ 
x_{B}\left( k\right)%
\end{array}%
\right) e^{-ink},
\end{align}%
with 
\begin{equation}
M\left( k\right) =-\left( t_{A}+t_{B}\right) I_{2}+\left( 
\begin{array}{cc}
0 & t_{A}+t_{B}e^{-ik} \\ 
t_{A}+t_{B}e^{-ik} & 0%
\end{array}%
\right) .
\end{equation}%
Here, $M\left( k\right) $ is identical to the SSH Hamiltonian up to a
constant term.

The dimerized Toda equation (\ref{TeleEq}) is reduced to a dimerized linear
wave equation for $\xi \rightarrow 0$,%
\begin{equation}
\frac{d^{2}}{d\tau ^{2}}\left( 
\begin{array}{c}
x_{A}\left( k\right) \\ 
x_{B}\left( k\right)%
\end{array}%
\right) =M\left( k\right) \left( 
\begin{array}{c}
x_{A}\left( k\right) \\ 
x_{B}\left( k\right)%
\end{array}%
\right) .  \label{EqB}
\end{equation}
Correspondingly, a nonlinear transmission line [Fig.\ref{FigIllust}(a)] is
reduced to a linear transmission line [Fig.\ref{FigIllust}(b)]. The
topological properties of this linear wave equation have been studied in a
transmission line\cite{QWalk} and also in acoustic systems\cite%
{Prodan,Lubensky,Nash,Sus,Kariyado,Takahashi,Mee,Abba,Mat}.

\textbf{Topological number}: The topological number for the\ SSH model is
defined by the chiral index,%
\begin{equation}
\Gamma =\int_{0}^{2\pi }\text{Tr}\left[ \sigma _{z}M^{-1}\left( k\right)
\partial _{k}M\left( k\right) \right] dk.  \label{TopoCharge}
\end{equation}%
Indeed, it is quantized, i.e., $\Gamma =1$ for $\left\vert t_{a}\right\vert
<\left\vert t_{b}\right\vert $ and $\Gamma =0$ for $\left\vert
t_{a}\right\vert >\left\vert t_{b}\right\vert $. Namely, the system is
topological for $\lambda <0$ and trivial for $\lambda >0$. We show the
eigenvalue of $M(k)$ for a finite chain in Fig.\ref{FigEdge}(a), where the
emergence of topological edge states is manifest for $\lambda <0$, as marked
in red. Characteristic topological properties are induced by topological
edge states.

\textbf{Dynamics}: The main issue of the present work is to inquire whether
the nonlinear system (\ref{TeleEq}) is also characterized by the topological
number (\ref{TopoCharge}). First of all, let us make the following
speculation. The left-hand side of Eq.(\ref{TeleEq}) is a monotonic smooth
function for $\xi $. It indicates that there is no phase transition in the
parameter $\xi $. Hence, we expect that the topological properties will be
inherited to the nonlinear system ($\xi \neq 0$) from the linear system ($%
\xi =0$).

In order to justify this speculation, we solve the nonlinear wave equation (%
\ref{TeleEq}) as well as the linear one ($\xi =0$) numerically under the
initial condition that we apply a nonzero voltage only at the left edge. We
present numerical results for the nonlinearity $\xi =0$, $0.1$, $1$, $5$ and 
$10$ in two typical phases at $\lambda =\mp 0.5$ in Fig.\ref{FigDynamics}.
In a topological phase ($\lambda =-0.5$), we observe a clear standing wave
displayed in red, which is due to the topological edge state. Although the
amplitude decreases as the nonlinearity $\xi $ increases, the overall
structure remains unchanged. 
On the other hand, 
there is no standing wave in a trivial phase ($\lambda =0.5$), irrespective
of the nonlinearity.

These properties are made clearer by making the Fourier analysis in
frequency $\omega $.\ We focus on the topological phase: See Fig.\ref%
{FigDynamics}(c1)$\sim $(c5). In the linear regime ($\xi =0$), there is a
single strong peak, which remains as it is in weak nonlinear regime ($\xi
=0.1$). In strong nonlinear regime ($\xi \geq 5$), many satelite peaks
emerge at the harmonic overtone frequencies. The intensity of the primal
peak decreases in strong nonlinear regime due to the emergence of the
satellite peaks. Such an effect of the nonlinearity manifests itself in the
Fourier decomposition. On the other hand, all peaks are quite tiny
irrespective of the linear or nonlinear regime and furthermore we do not see
overtone satellite peaks in the trivial phase, as in Fig.\ref{FigDynamics}(f1)%
$\sim $(f5).

\begin{figure*}[t]
\centerline{\includegraphics[width=0.98\textwidth]{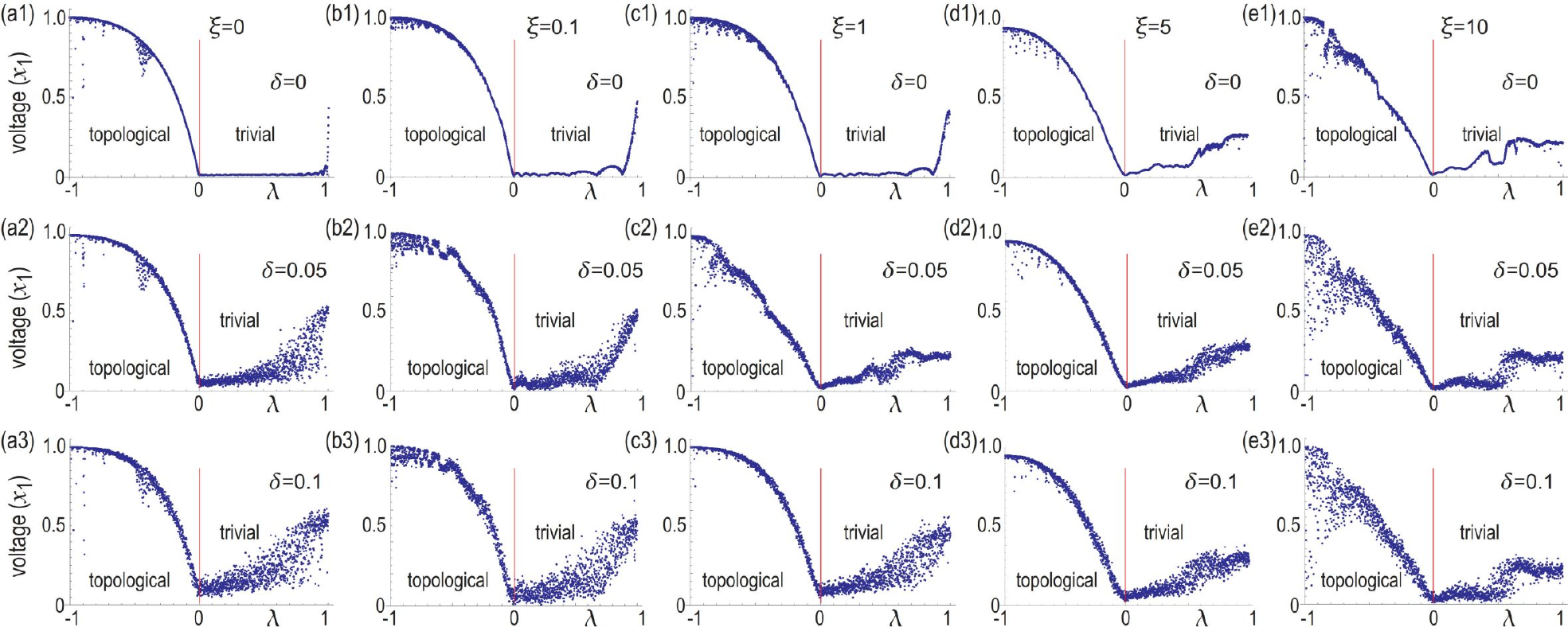}}
\caption{Amplitude of the dimensionless voltage $x_{1}$ at the left edge
after long enough time. The horizontal axis is $\protect\lambda $: The
system is topological (trivial) for $\protect\lambda <0$ ($\protect\lambda >0
$). (a1)$\sim $(a3) $\protect\xi =0$, (b1)$\sim $(b3) $\protect\xi =0.1$,
(c1)$\sim $(c3) $\protect\xi =1$, (d1)$\sim $(d3) $\protect\xi =5$, and (e1)$%
\sim $(e3) $\protect\xi =10$. 
(a1)$\sim $(a3) Pure system ($\protect\delta =0$); (a2)$\sim $(e2) 5\% randomness ($\protect\delta =0.05$); (a3)$%
\sim $ (e3) 10\% randomness in inductance ($\protect\delta =0.1$).}
\label{FigPhase}
\end{figure*}

These behaviors are actually independent of $\lambda $, although the
amplitude may change. To show it, we show the amplitude at the left edge as
a function of $\lambda $ in Fig.\ref{FigPhase}, where the topological phase
transition is clearly observed at $\lambda =0$ independent of nonlinearity $\xi $. 

\textbf{Bulk-edge correspondence:} These results are interpreted as follows.%
Once we start with a voltage localized initially at one
edge, there remains a finite voltage at the edge in the topological phase.
This is because the topological edge mode is almost isolated from all other
modes, although a certain amount of voltage propagates into the chain. The
amount of voltage lost from the edge becomes larger as the nonlinearity $\xi 
$ increases. On the other hand, since there is no localized edge mode in the
trivial phase, almost all amounts of voltage applied at the edge propagate
into the chain. These results are manifestation of the nonlinear bulk-edge
correspondence in physical observable quantities.

\textbf{Randomness effect:} We study how the topological dynamics is robust
against randomness in inductors. We have introduced randomness into
inductors uniformly distributing from $-\delta $ to $\delta $ by the
procedure $L\pm \ell \longmapsto \left( L\pm \ell \right) \left( 1+\eta
\delta \right) $, where $\eta $ is a random variable ranging from $-1$ to $1$%
. Numerical results are shown in Figs.\ref{FigEdge}, \ref{FigDynamics} and %
\ref{FigPhase} for some values of $\delta $. Clearly the dynamics is almost
identical between the pure system and disordered systems, which indicates
that the topological edge states are robust against randomness. On the other
hand, there is a new phenomenon due to randomness in the trivial phase,
where the amplitude becomes nonvanishing. This is because the voltage
propagation is localized due to the randomness, which is a reminiscence of
the Anderson localization\cite{Ander}.

Comments are in order. (i) The amplitude becomes large around $\lambda
\simeq 1$ in Fig.\ref{FigPhase}, although the system is trivial. In this
parameter region, the system is almost dimerized, where the voltage cannot
propagate into the bulk. (ii) The amplitude takes the minimum at $\lambda =0$%
\ irrespective of the nonlinearity $\xi $\ and the disorder $\delta $ in Fig.%
\ref{FigPhase}. This phenomenon would be due to the existence of a Toda
soliton at $\lambda =0$ for all $\xi $. The soliton moves with a constant
velocity, which means that the voltage at the edge rapidly decreases as a
function of time. Furthermore, the soliton dynamics is robust for disorders.
Hence, the topological phase transition is rigid for all $\xi $ and $\delta $%
.

\textbf{Discussion}: The topological analysis is familiar in condensed
matter physics and also in linear systems such as electric circuits and
acoustic systems. However, it is highly nontrivial in nonlinear systems,
where a general argument would be formidable. It would be necessary to make
individual study in various typical models.

In this work we have made a first step toward the problem by taking the
dimerized Toda lattice model, where we have identified a parameter
controlling nonlinearity. An important feature is that the model has an
adiabatic connectivity to a linear system governed by the SSH model. We have
carried out a numerical analysis to show that the bulk-edge correspondence
is inherited from the linear system to the nonlinear system. Indeed, we
have numerically checked in Figs.\ref{FigDynamics} and \ref{FigPhase} that
the topological dynamics in nonlinear systems is the same as the one in the
linear system. We may conclude that the topological number well captures the
topological dynamics in the dimerized Toda lattice. Our scheme will be
applicable to make topological analysis in other nonlinear systems.

The author is very much grateful to M. Kawamura, S. Katsumoto, Pierre
Delplace and N. Nagaosa for helpful discussions on the subject. This work is
supported by the Grants-in-Aid for Scientific Research from MEXT KAKENHI
(Grants No. JP17K05490 and No. JP18H03676). This work is also supported by
CREST, JST (JPMJCR16F1 and JPMJCR20T2).

\end{document}